\documentclass[useAMS,usenatbib]{mn2e}
\usepackage{graphicx}
\usepackage{amssymb}
\usepackage{amsmath}
\usepackage{times}
\usepackage{epsfig}

\title[Monthly Notices]
  {The Evolution of Radio Loud Active Galactic Nuclei as a Function of Black Hole Spin}
\author[D. Garofalo et al.]
  {D.~Garofalo,$^1$\thanks{david.a.garofalo@jpl.nasa.gov}
  D.A.~Evans,$^2$ R.M.~Sambruna,$^3$  \\
  $^1$Jet Propulsion Laboratory, California Institute of Technology,
      Pasadena, CA 91109, USA\\ $^2$Massachusetts Institute of
      Technology, Kavli Institute for Astrophysics and Space Research,
      77 Massachusetts Avenue, Cambridge, MA 02139,
      USA\\$^4$Astrophysics Science
      Division, Mail Code 662, NASA Goddard Space Flight Center,
      Greenbelt, MD 20771, USA} \date{Released 2010 Xxxxx XX}

\pagerange{\pageref{firstpage}--\pageref{lastpage}} \pubyear{2010}

\def\LaTeX{L\kern-.36em\raise.3ex\hbox{a}\kern-.15em
    T\kern-.1667em\lower.7ex\hbox{E}\kern-.125emX}

\begin{document}

\label{firstpage}

\maketitle

\begin{abstract}
Recent work on the engines of active galactic nuclei jets suggests
their power depends strongly and perhaps counter-intuitively on black
hole spin.  We explore the consequences of this on the radio-loud
population of active galactic nuclei and find that the time evolution
of the most powerful radio galaxies and radio-loud quasars fits into a
picture in which black hole spin varies from retrograde to prograde
with respect to the accreting material.  Unlike the current view,
according to which jet powers decrease in tandem with a global
downsizing effect, we argue for a drop in jet power resulting directly
from the paucity of retrograde accretion systems at lower redshift $z$
caused by a continuous history of accretion dating back to higher $z$.
In addition, the model provides simple interpretations for the basic
spectral features differentiating radio-loud and radio-quiet objects,
such as the presence or absence of disk reflection, broadened iron
lines and signatures of disk winds.  We also briefly describe our
models' interpretation of microquasar state transitions.  We highlight
our result that the most radio-loud and most radio-quiet objects both
harbor highly spinning black holes but in retrograde and prograde
configurations, respectively. 
\end{abstract}


\section{Introduction}

Over the past decade and a half, our understanding of the dynamics of
active galaxies (AGN) has made significant strides.  Once of marginal
interest astrophysically, black holes have taken center stage,
becoming an integral part of galactic dynamics and evolution.  The
picture that has emerged involves the presence of supermassive black
holes at the center of most if not all galaxies, with active galaxies
interacting with these black holes via accretion, producing winds and
jets.  There is overwhelming evidence that galaxies are interconnected
with black holes to the extent that the cosmic evolution of both is
coupled (Kormendy \& Richstone 1995; Magorrian et al 1998; Gebhardt et
al 2000; Ferrarese \& Merritt 2000; Tremaine et al 2002; Marconi \&
Hunt 2003).  Despite this, there are uncertainties in our
understanding of how black hole engines produce jets, how such
structures are collimated over large scales, and how galaxies that
harbor these powerful engines evolve over cosmic time.

The theoretical framework in which we have attempted to address such
questions over the past two decades involves the ``spin paradigm''
(Blandford 1990; Wilson \& Colbert, 1995; Moderski, Sikora \& Lasota,
1998), whereby high black hole spin can lead to powerful, radio-loud,
jetted AGN, while low black hole spin to radio-quiet, weak or
non-jetted AGN where jet power is related to black hole spin via the
Blandford-Znajek mechanism (Blandford \& Znajek 1977; henceforth BZ).
The literature on the BZ effect and black hole spin-related processes
for jet formation in relation to radio-loud AGN is extensive.  In its
most recent guise, the spin paradigm enters the literature in various
places and with differing results.  Using the BZ mechanism, Mangalam,
Gopal-Krishna \& Wiita (2009) propose a scenario leading to low black
hole spins in advection dominated accretion flows under the assumption
that the magnetic flux threading the black hole is insensitive to
spin.  Similar ideas grounded in general relativistic
magnetohydrodynamic simulations suggest that high spin for higher
order spin power dependence in thick accretion geometries vs thin
accretion flows, produce the conditions for the $3$ order of magnitude
difference in power between radio-loud and radio-quiet AGN
(Tchekhovskoy et al 2010).  In work that closely follows the results
of numerical simulations that supports the work of Nemmen et al (2007)
on high black hole spin in geometrically thick flows, Benson \& Babul
(2009), also using advection dominated thick accretion flow models,
arrive at a maximum spin of $0.92$ for black holes, suggesting that a
spin-equilibrium scenario is compatible with the linear accretion rate
dependence of jet power found in recent observations of AGN jets
(Allen et al 2006).

In addition to the question of the radio-loud/radio-quiet division in
active galaxies, a division within the group of jetted AGN is observed
and a classification is produced according to differences in power and
jet collimation (Fanaroff \& Riley 1974).  Attempts at understanding
this Fanaroff-Riley (FR) division have also been framed in the context
of the spin-paradigm in an attempt to explore the possibility of
differences in spin between FRII and FRI objects.  Meier (1999)
suggested a magnetic switch operating at high spin triggers the
production of relativistic jets.  Using the BZ effect and hybrid Meier
model, Daly (2009b) calculated the cosmological evolution of black
hole spins, showing that they are largest (about unity) for the FRII
sources at high redshift and decrease (to about 0.7) for radio-loud
sources at lower redshift.  The idea that the engine of jets
determines their FR morphology spans the works from Rees (1982) to
Hardcastle et al (2007).  A theoretical approach to determining black
hole spin that is model-independent, but assumes that spin changes
only by extraction of the reducible black hole mass, applied to a
small subset of powerful radio galaxies, finds that they harbor low
spinning black holes (Daly 2009a).  This work suggests also that the
FRI/FRII division is due to environment.  Such environmental effects
on jet morphology, unrelated at least directly to the properties of
the central engine of the jet, have been addressed starting with De
Young (1993) and more recently with Gopal-Krishna and Wiita
(2000). Black hole mass dependence (Ghisellini \& Celotti 2001) and
accretion dependence of the FRI/FRII division have also been explored
(Marchesini et al 2004).

Under the assumption that jets are produced by the combined effort of
the BZ and Blandford-Payne (Blandford \& Payne 1982; henceforth BP)
effects, recent numerical studies of general relativistic
magnetohydrodynamics of black hole accretion flows suggest a tight
link between jet power and black hole spin (Garofalo, 2009a, 2009b).
Whereas BZ involves jets produced via spin-energy extraction from the
black hole, BP is a jet mechanism that originates in the accretion
disk via mass-loading of disk gas onto large scale magnetic fields.
The black hole spin dependencies of these processes have direct
implications for the jets produced in AGN.  According to these
studies, largest BZ power occurs for highly retrograde accretion
systems with respect to the black hole and less so for prograde ones,
transitioning to zero power at zero spin while the BP power is also
maximized for highly retrograde accretion but monotonically decreases
toward high prograde ones.

In the context of this framework, we attempt in this paper to
construct a physical foundation for the morphology and evolution of
jet-producing AGN.  At the heart of our picture lies the notion that
accretion onto spinning supermassive black holes tends, in time, to
produce increasingly prograde accretion systems, in which lower jet
output is hosted by more stable accretion configurations.  In terms of
the radio-loud AGN population, ``high excitation radio galaxies'',
evolve over cosmic timescales toward ``low-excitation radio
galaxies'', as black holes conspire with their host galaxy in evolving
toward prograde accretion states from an earlier phase of retrograde
accretion flow.  In addition, we suggest a retrograde vs prograde
division for the engines of radio-loud objects whose interaction with
the external environment produces the distribution of the Fanaroff \&
Riley classification, according to which the powerful, highly
collimated jets of the FRII class, are retrograde spin systems while
the mostly less powerful, less collimated FRI sources embedded in
gas-rich environments are the late-state evolution toward prograde
accretion systems.  Our interpretation of the radio-loud/radio-quiet
division will be that the most radio-loud and the most radio-quiet AGN
both harbor rapidly rotating black holes with their accretion angular
momentum vectors determining the difference, while the FRI/FRII
division in our model is based on a combination of nuclear properties
and environment. In addition to this, our model unifies radio-loud,
radio-quiet, and FRI, FRII objects, to the extent, as we will
illustrate, that radio-loud objects of the FRII class evolve either
into radio-loud FRI objects or into radio-quiet AGN. We point out,
furthermore, that the theoretical frameworks grounded in the
spin-paradigm discussed above, all suffer from a ``spin paradox''
(David L. Meier) to be discussed in a follow-up paper in which we
illustrate its resolution within our framework (Meier \& Garofalo, in
preparation).  In section \ref{observation} we highlight the relevant
observed properties of radio-loud AGN which constitute the pieces of
the theoretical puzzle we construct in Section \ref{Theory}.  Section
\ref{Conclusions} briefly addresses the radio-loud/radio-quiet
dichotomy in this model and its extension to microquasars and then
concludes.

\section{Radio-loud AGN: Observational picture}

In this section we emphasize the main observational features
concerning the radio-loud population and give a broad-brush
description of past attempts and difficulties in combining such
features into a theoretical framework.
\label{observation}

\subsection{The Excitation Dichotomy in Radio-Loud AGN}

Extensive optical and X-ray surveys show clear evidence for a
fundamental dichotomy in the properties of radio-loud AGN, which is
directly related to the mode of accretion onto the central
supermassive black hole.

``High-excitation radio galaxies'' (HERGs), those with prominent
emission lines in their optical spectra, have standard, geometrically
thin accretion disks and accrete at a significant fraction of their
Eddington limits. These sources are heavily obscured in the X-ray by
columns in excess of $10^{23}$ cm$^{-2}$, consistent with AGN
unification (Donato et al. 2004; Evans et al. 2006). HERGs show
weaker, if any, neutral, inner-disk, broadened Compton reflection
continuua compared to radio-quiet AGN (Reeves \& Turner 2000; Grandi
et al. 2002). Furthermore, most of these sources have narrow,
typically unresolved neutral Fe K$\alpha$ lines (Evans et al. 2004,
2006), which indicates that the primary X-ray emission is being
reprocessed far from the inner disk regions. HERGs tend to inhabit
isolated environments, at least at low redshift, and often show
evidence for recent mergers, consistent with the idea that they derive
their power from the accretion of cold gas (Hardcastle, Evans, \&
Croston 2007).  {\it Spitzer} IRS spectroscopy (Ogle et
al. 2006) shows that HERGs are luminous MIR emitters, and the
detection of strong 9.4$\mu$m silicate absorption implies that they
possess molecular tori.

On the other hand, ``low-excitation radio galaxies'' (LERGs) -- those
with few or no observed optical emission lines -- lack any of the
features required by standard AGN unification models. Their X-ray
emission is dominated by a parsec scale jet, they have radiatively
inefficient accretion flows ($L/L_{Edd}\sim10^{-(5-7)}$ - where
$L$ is luminosity and $L_{Edd}$ is the Eddington luminosity), and
they show no evidence at all for an obscuring torus (Hardcastle,
Evans, \& Croston 2006; Ogle et al 2006). These sources tend to
inhabit hot, gas-rich environments, such as groups and clusters. It
has been suggested by Hardcastle, Evans, \& Croston (2007) that the
jet outbursts in LERGs derive their power from the Bondi accretion of
this IGM/ICM gas. This result holds for the majority of LERGs, but
cannot be applied to the most powerful jets in clusters of galaxies,
such as MS0735.6+7421 (McNamara et al. 2009), whose kinetic power
exceeds that available from Bondi accretion.

Finally, while LERGs show no evidence of ionized outflows of gas or
winds from their central regions, evidence is emerging for such
outflows in HERGs, at least in X-rays (Reeves et al 2009; Tombesi et
al submitted), although less so than for their radio-quiet
counterparts.  


\begin{table}
\begin{tabular}{p{2.5cm}p{2cm}p{2cm}}
\hline
 & Low-excitation (LERG) & High-excitation (HERG) \\
\hline
Definition & No narrow optical line emission. & Prominent optical emission lines, either narrow (NLRG) or broad (BLRG), or quasar. \\ \hline

Fanaroff-Riley classification & Almost all FRIs are LERGs at $z$=0.  Significant population of FRIIs at $z \sim$0.5. & Most FRIIs are HERGs, as are a handful of FRIs (e.g., Cen~A). \\ \hline

X-ray spectra & Jet-related unabsorbed power law only. Upper limits only to 'hidden' accretion-related emission. & Jet-related unabsorbed power law + significant accretion contribution (heavily absorbed in NLRGs). \\ \hline

Accretion-flow type & Highly sub-Eddington. Likely radiatively inefficient. & Reasonable fraction of Eddington. Likely standard accretion disk. \\ \hline

Optical constraints & Strong radio/optical/soft X-ray correlations. Optical emission is jet-related. & Strong radio/optical/soft X-ray correlations. Optical emission is jet-related. \\ \hline



\hline
\end{tabular}
\label{summary}
\caption{Overview of the properties of low- and high-excitation radio galaxies}
\end{table}


\subsection{The Fanaroff-Riley Dichotomy in LERGs and HERGs and its Redshift Dependence}

HERGs and LERGs can be further classified according to whether they
belong to the Fanaroff \& Riley classification FRI or FRII (Fanaroff
\& Riley 1974), whose differences are related to the power of the jet
and the physics of mass-loading (entrainment).  FRI-type radio
galaxies display ``edge-darkened'' radio morphology, with generally
weak jets that are poorly collimated on kpc-scales. FRIIs are
typically more powerful and more collimated. Whereas most FRIIs are
HERGs and most FRIs are LERGs, {\it Chandra} observations indicate the
existence of mixed FRII LERG and FRI HERG states.  FRI HERGs are very
rare.  One example is the nearest AGN, Centaurus A (Evans et
al. 2004). On the other hand, there is a significant population of
FRII LERGs, which occur at what is arguably an intermediate redshift
($z=0.5-1$) for these objects, and essentially zero cases at low
redshift ($z < 0.1$).  FRII LERGs tend to lie in gas-rich groups or
cluster-scale environments.

The appearance of radio-loudness in the forms discussed above and the
connection to radio-quiet AGN forms the subject of the following
sections.  Combinations of radiatively inefficient/efficient accretion
flow with high/low spinning black holes (the spin paradigm) are only
partially successful in modeling the observations.  The observed
weakness of disk winds and broadened reflection features in FRII HERGs
compared to radio-quiet objects remains unexplained.  Recent work
points to a possible resolution of such problems by suggesting that
radiative efficiency in very high accretion rate systems may be
accompanied by fully ionized inner regions, thereby explaining the
weakness of the reflection component (Sambruna et al 2009 and
references therein).  Explaining LERG jet systems in gas-rich clusters
and the energetics of the FRII class is also problematic as mentioned
above.  For LERG jets, the observed linear relation between jet power
and accretion power (Allen et al 2006) is expected assuming the BZ
effect is operating but energetically can only be explained if the
black holes are maximally spinning (Nemmen et al 2007; Benson \& Babul
2009), which, in turn, implies there is fine-tuning in the spin
parameter.  The attempt to extend this linear relation between BZ jet
power and accretion rate fails for the FRIIs because the observed
accretion power is too weak.  And finally, models suggesting high spin
leads to powerful jets are statistically incompatible with mounting
evidence for high-spinning black holes in Seyfert galaxies as it
becomes problematic to assume such objects are all currently living
within their radio-quiet phase (Wilms et al 2001; Brenneman \&
Reynolds 2006; Fabian et al 2009; Zoghbi et al 2010).

In the next section we attempt to produce a black hole engine-based
theoretical framework that addresses these observed properties.  We
will argue that misaligned, or retrograde systems, in which disk gas
rotates opposite to that of the black hole, create the conditions for
powerful jets albeit in dynamically unstable configurations, whose
time evolution is toward energetically more stable conditions in which
the black hole and disk coexist in rotationally aligned states.  We
will suggest that in addition to radiative efficiency or lack thereof
and black hole spin, alignment vs misalignment between the angular
momentum of the black hole and the disk, is crucial.

\section{Radio loud AGN: Theoretical framework}

In this section we describe the theoretical framework within which we
fit the observational elements of the previous section. The
fundamental distinguishing feature of the theory (Garofalo, 2009a)
involves the change in size of the gap region that exists between the
inner edge of accretion disks located near the innermost stable
circular orbit (ISCO) and the black hole horizon.  The size of the gap
region changes as a function of black hole spin because the ISCO
depends on spin.  We quantitatively determine the effect that the
change in size of the gap region has on the BZ mechanism, on the BP
mechanism, and on the overall effect of the gap region on jet power
and collimation under the assumption that BZ and BP are the dominant
and mutually dependent mechanisms involved in jet production.
Finally, we include the effect of the gap region on accretion
efficiency.  We should emphasize that the fundamental feature of our
model involves the inverse relationship between accretion efficiency
and jet efficiency, and that section \ref{gap_paradigm} is focused on
illustrating the dependence of such efficiencies on the size of the
gap region.
\label{Theory}

\subsection{BZ and BP: quantitative approach}

We assume the spacetime is that of a rotating black hole and proceed
in Boyer-Lindquist coordinates for which the Kerr metric takes the
form,
\begin{eqnarray}
  dS^{2}&=&-\left(1-\frac{2Mr}{\rho^{2}}\right)
  dt^{2}-\frac{4Mar\sin^{2}\theta}{\rho^{2}}dt\,d\phi\\
  &+&\frac{\Sigma}{\rho^{2}}\sin^{2}\theta\, \nonumber
  d\phi^{2}+\frac{\rho^{2}}{\Delta}dr^{2}+\rho^{2}d\theta^{2},
\end{eqnarray}
where $M$ is black hole mass, $a$ is the dimensionless spin
parameter,
\begin{equation}
\rho^{2} = r^{2} + a^{2}\cos^{2}\theta,
\end{equation}
\begin{equation}
\Delta = r^{2}-2Mr+a^{2},
\end{equation}
and
\begin{equation}
\Sigma = (r^{2}+a^{2})^{2}-a^{2}\Delta\sin^{2}\theta.
\end{equation}

The basic equation describing the evolution of the large-scale
magnetic field within a Novikov \& Thorne (1973) acccretion disk is
obtained by following the relativistic analogue (Garofalo 2009b) of
the non-relativistic treatment of Reynolds et al (2006), by combining
Maxwell's equation
\begin{equation}
\triangledown_{b}F^{ab} = \mu J^{a},
\end{equation}
with a simplified Ohm's law
\begin{equation}
\label{Ohm}
J^{a} = \sigma F^{ab}u_{b},
\end{equation}
where $F^{ab}$ is the standard Faraday tensor, $\mu$ is the
permeability of the plasma, $J^a$ the 4-current, $u^a$ the 4-velocity
of the accretion disk flow, and $\sigma$ the effective conductivity
of the turbulent plasma.  This gives
\begin{equation}
\triangledown_{b}F^{ab} = \frac {1}{\eta} F^{ab}u_{b},
\label{disk}
\end{equation}
where $\eta = 1/\mu\sigma$ is the effective magnetic diffusivity.  The
equations are cast in terms of the vector potential, which is related
to the Faraday tensor via
\begin{equation}
 F_{ab}=A_{b,a}-A_{a,b},
\label{vec_pot}
\end{equation}
and, in particular, in terms of the component $A_{\phi}$ in the
coordinate basis of the Boyer-Lindquist coordinates.   

Ultimately, to examine BZ and BP powers, we need to derive the
magnetic flux and angle of the flux contours threading a hoop placed
at a given radius $r$.  The magnetic flux function is related to the
vector potential via Stokes' Theorem applied to the Faraday tensor
\begin{equation}
\psi \equiv \int_{S}F=\int_{S}dA=\int_{\partial{S}}A=2 \pi A_{\phi}, 
\end{equation}
where $S$ is a space-like surface with boundary $\partial S$
consisting of a ring defined by $r={\rm constant}$, $\theta={\rm
constant}$, and $t={\rm constant}$.  Since $A_{b}$ is specified up to
the gradient of a scalar function $\Gamma$,
\begin{equation}
A_{b}^{'} = A_{b} + \triangledown _{b} \Gamma,
\end{equation}  
the assumption of time-independence and axisymmetry gives us
\begin{equation}
A^{'}_{t} = A_{t}
\end{equation}
and
\begin{equation}
A_{\phi}^{'} = A_{\phi}.
\end{equation}
Thus, we need not specify the gauge uniquely beyond the statement of t
and $\phi$ independence.

The region outside the black hole and accretion disk is modeled as
force-free, satisfying
\begin{equation}
F^{ab}J_{b} = 0
\label{force-free}
\end{equation}
and
\begin{equation}
\triangledown_{b}F^{ab} =  \mu J^{a}.
\label{magnetosphere1}
\end{equation}
Outside the accretion disk we also impose the ideal MHD condition
\begin{equation}
F^{ab}u_{b} = 0,
\label{ideal}
\end{equation}
where $u^b$ is the 4-velocity of the (tenuous) plasma in the
magnetosphere and is determined by the condition that field lines
rigidly rotate.  The numerical details can be found in Garofalo 2009b.

We now consider the BZ and BP powers that result from the numerical
solution of the above equations under the assumption that the gap
region is not threaded by magnetic flux (i.e. the Reynolds conjecture
of a zero-flux boundary condition - Reynolds et al 2006; Garofalo
2009b), where this ability of the gap region to enhance magnetic flux
on the black hole is supported also by general relativistic MHD
simulations (McKinney \& Gammie, 2004).  The foundation of the
Reynolds conjecture stems from noting that within the gap region,
circular orbits are no longer stable and the accretion flow plunges
into the black hole.  In Reynolds et al (2006), it is argued that the
inertial forces within the gap region prevent magnetic flux that is
threading the black hole from expanding back into the disk.  Accretion
of magnetic field can result in a strong flux-bundle threading the
black hole, confined in the disk plane by the gap region.  We start by
evaluating the horizon-threading magnetic field as measured by ZAMO
observers from the flux values we obtain,
\begin{equation}
\label{zamo_b}
B_{H}=\sqrt{g_{11}}B^{r}
\end{equation}
with 
\begin{equation}
B^{r} = \ast F^{rb}u_{b},
\end{equation}    
where $\ast F^{ab}$ is the dual Faraday tensor and $u^{b}$ is the
four-velocity of the ZAMO observers evaluated in the equatorial plane
on the stretched horizon or membrane in the sense of the Membrane
Paradigm (Thorne et al 1986), and from this magnetic field value we
determine BZ power as (Thorne et al 1986),

\begin{equation}
L_{BZ} = 2 \times 10^{47}ergs \;
s^{-1}(\frac{B_{H}}{10^{5}G})^{2}m_{9}^{2}j^{2}
\end{equation}

where $B_{H}$ is the poloidal magnetic field threading the black hole, $j$
is the normalized angular momentum of the black hole or $a/M$, and $m_{9}$
is the black hole mass in units of $10^{9}$ solar masses.

\begin{figure*}
\centerline{\includegraphics[angle=-0,scale=0.60]{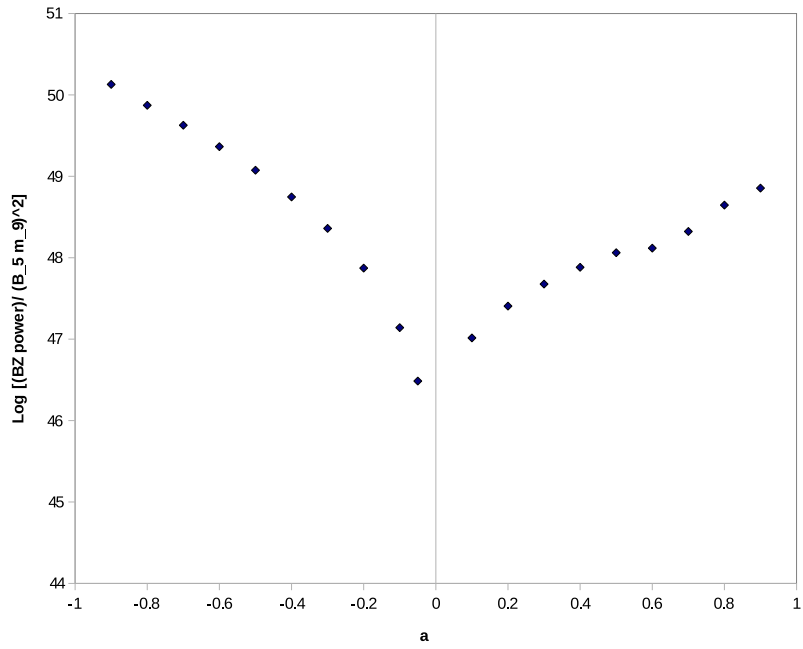}}
\caption{The BZ power vs spin accretion state (negative for retrograde
and positive for prograde).  }
\label{L_BZ}
\end{figure*}
\begin{figure*}
\centerline{\includegraphics[angle=-0,scale=0.60]{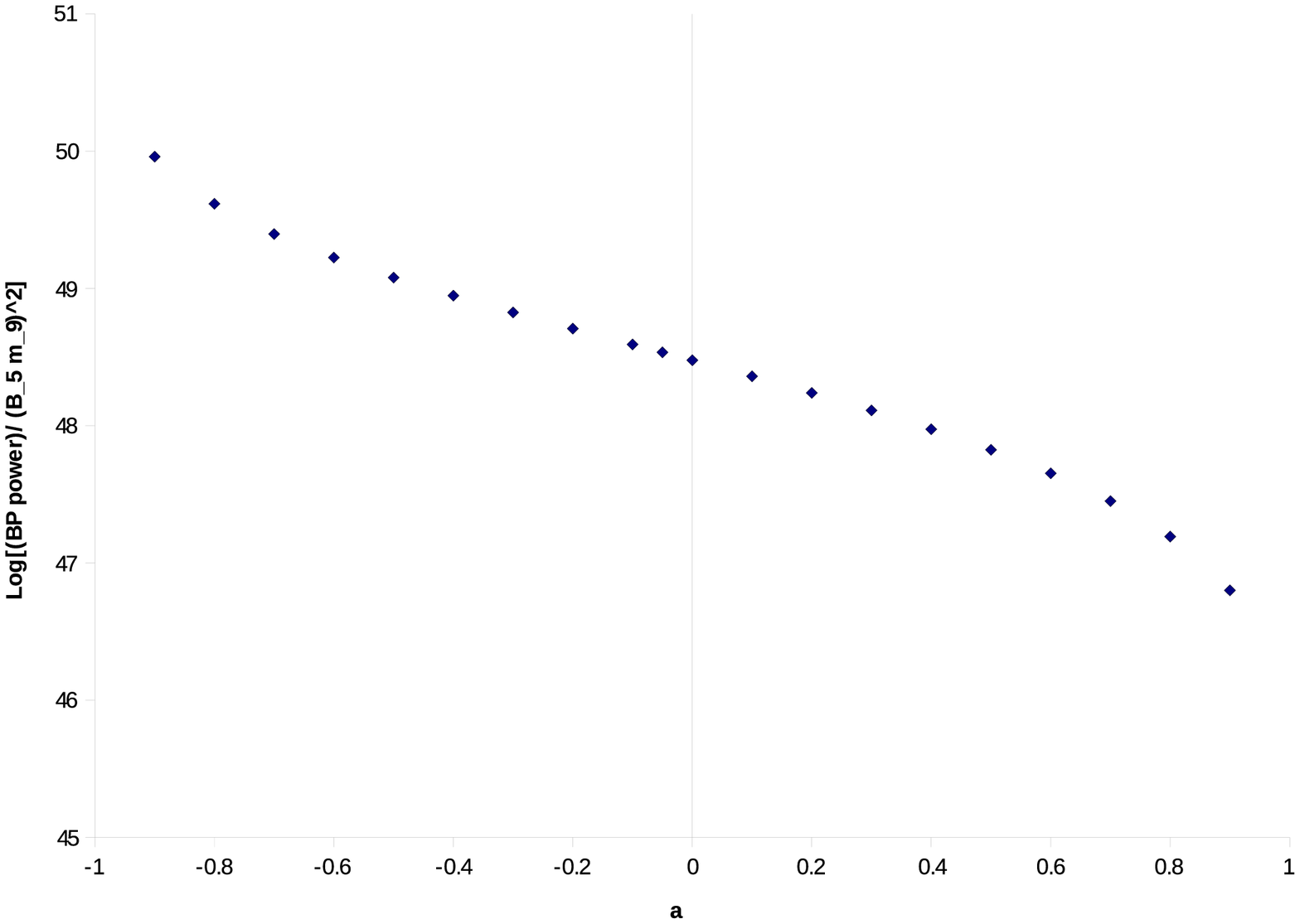}}
\caption{The BP power vs spin accretion state (negative for retrograde
and positive for prograde). 
}\label{L_BP}
\end{figure*}

BP power is similarly constructed but depends on magnetic field
strength threading the accretion disk, the bend angle with which the
magnetic field presents itself to the disk surface, or more precisely,
the radial extent over which the bend angle is sufficient for
mass-loading, and the Keplerian rotation rate which is also
spin-dependent.  Following Cao (2003) we have  

\begin{equation}
L_{BP} = \int B_{d}^{2} r^{2} \Omega dr 
\end{equation}

where $B_{d}$ is the field strength threading the disk and $\Omega$ is
the rotation of the disk magnetic field.  The radial extent over which
the bend angle is large enough for mass-loading, increases with
increase of spin in the retrograde regime and comes from the numerical
solution (Garofalo 2009b).  We show the power in BZ in Figure
\ref{L_BZ}, that in BP in Figure \ref{L_BP} and the overall power in
Figure \ref{Ledlow} assuming they combine as follows, where the
functions $\alpha$ and $\beta$ capture the effects on BP and BZ powers
from the numerical solution in the context of the Reynolds conjecture
(Garofalo 2009b).

\begin{equation}
L_{jet} = 2 \times 10^{47}ergs \; s^{-1}
\alpha\beta^{2}(\frac{B_{d}}{10^{5}G})^{2}m_{9}^{2}j^{2}
\label{jet_power}
\end{equation}

where

\begin{equation}
\alpha = \delta(\frac{3}{2}-j)
\end{equation}

and 

\begin{equation}
\beta = -\frac{3}{2}j^{3} + 12j^{2} - 10j + 7 -
\frac{0.002}{(j-0.65)^{2}} +\frac{0.1}{(j+0.95)}
+\frac{0.002}{(j-0.055)^{2}}.
\end{equation}
While $j$ spans negative values for retrograde spin and positive
values for prograde spin, a conservative value for $\delta$ of about
$2.5$ is adopted but our ignorance of how jets couple the BZ and BP
components restrict our ability to specify it and suggest that it
might well be larger by an order of magnitude or more.

While $\alpha$ can be thought of as the parameter that determines the
effectiveness of the BP jet as a function of spin, $\beta$ captures
the enhancement on the black hole of the disk-threading field, both
within the context of the Reynolds conjecture.  The parameter $\delta$
determines the effective contribution of BP to overall jet power, with
larger values shifting jet power efficiency more toward the retrograde
regime.  Because, as we point out in the next section, our focus is on
the collimating properties of BP jets, we have been conservative in
choosing a small $\delta$ of order unity.  It is important to point
out that the no-flux boundary condition's effect is to increase the BZ
and BP efficiency toward the retrograde regime.  The consequence of
this for prograde spin is to produce a flattening of the overall jet
power so that low prograde black hole spins have powers that increase.
The behavior of the power in Figure \ref{Ledlow} follows from the
assumption of a force-free magnetosphere but changes if this
assumption is relaxed.  A non-negligible inertia for fast rotating
magnetospheres and the added centrifugal barrier that would ensue,
would make it more difficult for plasma to be advected inward through
the diffusionless gap region, a situation that arises more for the
high prograde regime as the rotation is greater.  As a result, a more
realistic magnetosphere should further shift the overall jet power
efficiency toward the retrograde regime, thereby further flattening
the spin dependence in the prograde one.

\subsection{Jets vs. accretion: the gap paradigm}

We combine these results and produce the following illustration of the
combined effect of BZ and BP on jet power and collimation.

\begin{figure*}
\centerline{\includegraphics[angle=-0,scale=0.3]{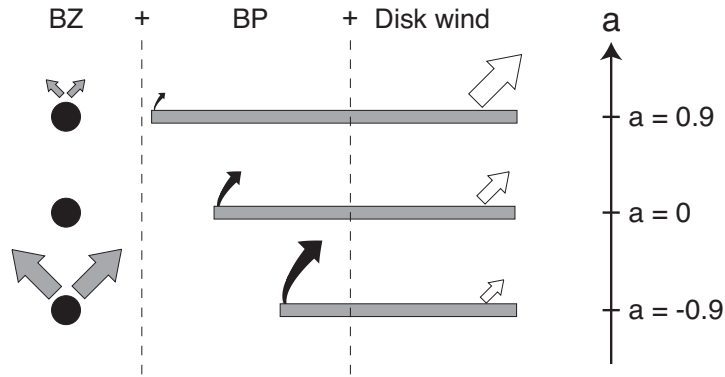}}
\caption{Left column: Despite a small gap region, high prograde
spinning systems (upper panel) still produce BZ jets because the spin
is large, but they are weaker than for high retrograde systems where
the gap region is larger (lower panel).  Center column: The increased
size of the gap region as the black hole spin goes from high prograde
(upper panel) toward high retrograde (lower panel) makes the BP jets
stronger as indicated by the length of the arrows originating in the
accretion disk.  However, as discussed in Section \ref{Conclusions},
this BP efficiency may become negligible at high prograde spin
(Neilsen \& Lee 2009).  Right column: The increased size of the gap
region for higher retrograde systems produces weaker disk winds due to
the decrease in the amount of reprocessable gravitational potential
energy near the black hole. }\label{BZ}
\end{figure*}

\begin{enumerate}

\item In the left column in Figure \ref{BZ} we show the difference
between accretion systems around spinning black holes as a function of
the spin parameter and type of accretion (i.e. prograde
vs. retrograde) and its effect on the BZ power.  We use negative spin
values to indicate retrograde accretion while positive spin values
indicate prograde accretion.  Note how the gap region is larger as the
spin increases from large prograde to large retrograde.  The shrinking
of the gap region in the prograde direction, produces an asymmetry in
BZ power that favors retrograde systems.  As a result, retrograde BZ
power is larger as indicated by the length of the arrows.  The basic
physics behind this power dependence on spin is that larger gap region
allows greater magnetic flux accumulation on the black hole (Garofalo,
2009b).

\item In the center column of Figure \ref{BZ} we present a schematic
of the monotonic dependence of BP on spin by using arrows that
originate in the disk (since BP jets are disk-launched).  The length
again indicates the magnitude of the jet and we highlight how the
power increases from high prograde to high retrograde.  Here, as well,
the size of the gap region is instrumental in that larger gap regions
produce greater magnetic flux on the black hole, which leads to
greater bending of inner-disk-threading magnetic field lines, which,
in turn, leads to greater mass outflow and more effective BP jets
(Garofalo 2009a).

\item By considering both the left and central columns of Figure
\ref{BZ} we combine the effects and produce an overall picture for the
jets in retrograde vs prograde black hole accretion systems.  Assuming
that BP jets originating in the disk/corona are essential for jet
collimation (McKinney \& Narayan 2007; Bogovalov \& Tsinganos 2005;
Meier et al 2001), and that such collimation is directly related to
jet acceleration (Tchekhovskoy, McKinney \& Narayan, 2009), the
strongest and most collimated jets occur for high retrograde systems
while high prograde systems produce low-power, weakly collimated jets
due to the weakness of BP, despite the presence of strong BZ.  Our
heuristic, qualitative scenario proposes that the jet itself comes
from the BP effect but is sparked, in some sense, by the BZ mechanism.
It is worth pointing out that we are not illustrating a precise
mechanism for producing jets, but simply highlighting the
compatibility that arises between theory and observation if BP and BZ
conspire as described above.  In addition, surviving MHD disruptions
such as the kink mode instability (Nakamura \& Meier 2004), appears to
occur in combination with the presence of a disk jet and relativistic
bulk motion (Hardee \& Hughes, 2003; McKinney \& Blandford, 2009) The
diagram, therefore, illustrates these ideas with uncollimated jets for
the high prograde case resulting from small BP, no jet in the zero
spin case, and a powerful, collimated jet for the high retrograde case
resulting from strong BP.
\end{enumerate}

We highlight two aspects of the BZ/BP scenario.  The first is that
whereas jet power displays a roughly flat spin dependence for a range
of prograde spin values, this does not occur for retrograde spin.
This indicates that unlike for retrograde spins, there is a range of
prograde spin values for which jet power is weakly dependent on spin.
We also re-emphasize the fact that BP jets do not simply add to the
overall jet power according to our picture; they serve the fundamental
purpose of collimation (a feature motivated in the diagrams of Figure
\ref{BZ}).

In addition to the effect that the location of the disk inner edge has
on jet production, the size of the gap region is also connected to the
energetics of the disk itself.  Recent work extending standard,
radiatively efficient accretion disk theory to self-consistently
include the effects of magnetized coronae and disk winds, shows that
larger gap regions inhibit or limit the presence of disk winds (Kuncic
\& Bicknell 2004, 2007).  The further out in radial position for the
disk inner edge, the less gravitational power is available to be
reprocessed in the disk to produce mass outflows further out in the
disk.  This is illustrated in the right column of Figure \ref{BZ}.
With only these considerations on the spin-dependent size of the gap
region, we introduce the predictions of our model for the cosmological
evolution of AGN (S.\ref{cosmo_evolution}), and its implications for
the Fanaroff-Riley dichotomy (S.\ref{Fanaroff-Riley}) and the observed
X-ray characteristics of radio-loud AGN (S.\ref{Fe_alpha}).

\label{gap_paradigm}

\subsection{Cosmological Evolution of Radio-Loud AGN}

Our story focuses on the fraction of objects that involve retrograde
accretion onto rapidly spinning black holes, most likely formed in
mergers of equal mass black holes or in mergers of black holes where
the larger one spins rapidly and the merger with the smaller one is in
the prograde direction (Hughes \& Blandford 2003).  As long as the
angular momenta of the disk, $J_{d}$, and that of the black hole,
$J_{h}$, satisfy
\begin{equation}
cos\theta < -\frac {J_{d}}{2J_{h}},
\end{equation}
where $\theta$ is the angle between the two angular momentum vectors
(King et al. 2005), counteralignment between disk and black hole
occurs (i.e. $\theta=\pi$) and a retrograde accretion state is formed.
Because the merger produces a gas-rich environment mainly in the form
of cool, molecular gas, that feeds the black hole at relatively high
accretion rates (Barnes \& Hernquist 1991), we associate this initial
phase with HERGs.  In other words, the relatively high accretion rates
are such that the accretion flow is close to a standard, radiatively
efficient Novikov \& Thorne disk.  The powerful, highly collimated
jet, on the other hand, is a direct consequence of the fact that the
system is in a highly retrograde accretion state (i.e. the black hole
engine is operating at maximum efficiency due to the presence of
strongest BZ and BP effects).  In other words, these FRII HERGs are
powered by nuclear engines in the lower panel of Figure \ref{BZ}.

\begin{figure*}
\centerline{\includegraphics[angle=-0,scale=0.3]{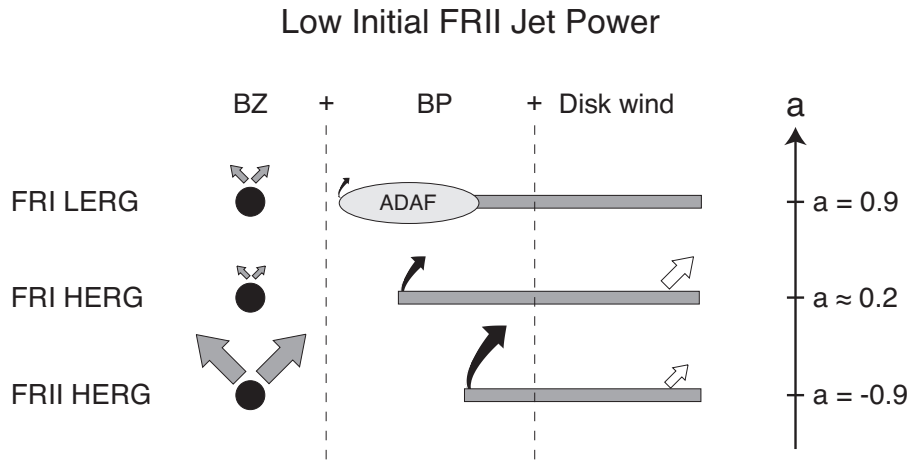}}
\caption{The evolution in time of lower power FRIIs, less able to
expel the cold gas, thereby accreting in the cold gas phase for
prolonged periods and thus into the prograde spin regime as HERGs.
Time increases upward.  }\label{BP}
\end{figure*}

We now consider the evolutionary paths of two initial FRII HERGs whose
jet powers differ, due perhaps to different ratios of black hole mass
to the amount of post-merger cold gas.  Recent work indicates that
high redshift, $z\sim2$, radio-loud quasars, can deliver $\sim10\%$ of
the jet energy to the ISM, sufficient to expel the cold gas via
outflows up to 1000 $km/s$ in $\sim10^{7}yrs.$ (Nesvadba et al. 2008).
If the loss of cold gas due to this radio-mode AGN feedback leads to a
drop in the pressure and density of the ISM, it is expected that
accretion will switch to hot Bondi-fed advection-dominated accretion
flow (ADAF - Narayan \& Yi 1995) from cold, thin-disk accretion at
higher redshift as galaxies expand to $\sim3$ times their size
(Mangalam, Gopal-Krishna \& Wiita 2009 and possibly as envisioned in
Antonuccio-Delogu \& Silk 2010).  If the initial FRII HERG is of lower
jet power (Fig. \ref{BP}), the expulsion of cold gas is less
effective, and the transition to hot Bondi-fed ADAF accretion is
slower.  In this case, the initial FRII HERG transitions to an FRI
HERG.  The FRI nature of the system originates in the fact of prograde
accretion as mentioned in the Introduction, while the HERG label
corresponds to the relatively high accretion rate and thus to
radiatively efficient accretion flow.  In other words, the spin-up
toward the prograde direction occurs faster than the transition from
radiatively efficient accretion to ADAF accretion.

\begin{figure*}
\centerline{\includegraphics[angle=-0,scale=0.3]{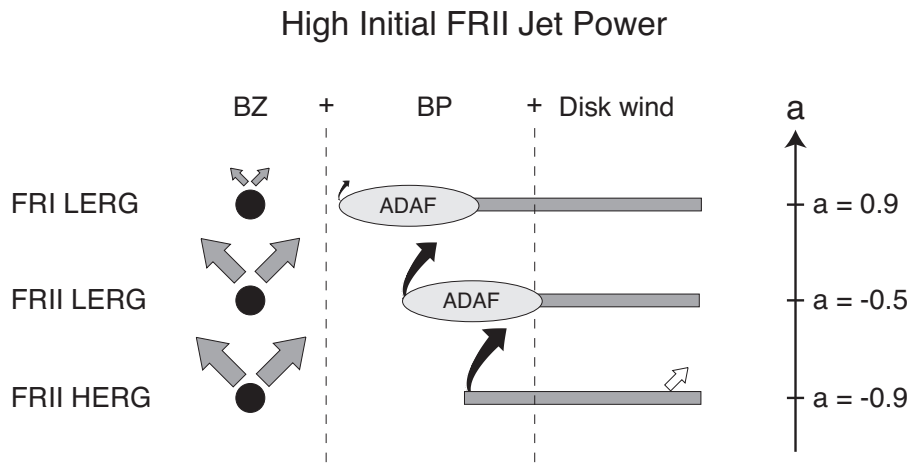}}
\caption{The evolution in time of higher power FRIIs, more able to
expel the cold gas, thereby accreting already in the ADAF phase while
still in the retrograde regime.  Time increases upward.
}\label{BZ+BP}
\end{figure*}

If, on the other hand, the initial FRII HERG jet is more powerful and
more effective in expelling the cold gas, the initial FRII HERG
transitions to an FRII LERG, due to the fact that the system remains
retrograde but the accreting gas is now hot and enters the ADAF phase
(i.e. the black hole has not accreted enough to have been spun down
and then up again in the prograde regime-Fig. \ref{BZ+BP}).

There are two basic processes working together here.  The first
involves accretion feeding a black hole in a retrograde state.  {\em
Continued accretion initially spins the black hole down toward zero
spin and then up again in a prograde state where the angular momentum
vector of the black hole is parallel to that of the accretion flow.}
The other process involves the accretion state itself.  If the
accreting gas is cold, the accretion flow radiates in an efficient
manner.  However, as described, the FRII HERG jets expel cold gas and
the timescales on which they accomplish this depends on jet power.
Therefore, more or less powerful FRII HERG, cold gas-expelling jets,
create the conditions that produce a mixing of FRIIs and FRIs with the
HERG and LERG states.  Nevertheless, continued cosmological evolution
of both FRII LERGs and FRI HERGs inevitably leads toward FRI LERG
states, so the mixed states appear at intermediate redshifts as is
observed (Hardcastle, Evans \& Croston, 2006). In other words, the
appearance of these transitional states between FRIIs and FRIs is
observationally in agreement with our model in that mixed states are
sandwiched between FRII HERGs and FRI LERGs.  From a general
perspective, cosmological evolution in this model produces FRI LERGs
from FRII HERGs.  As a direct consequence of this, LERGs have higher
black hole masses than HERGs, as is observed (Smolcic et al. 2009;
Smolcic 2009).

The boundary between FRIIs and FRIs produces other interesting
morphologies.  As a consequence of the fact that BZ power drops to
zero at zero black hole spin, there must be a transition period in
which the jet engine turns off.  The timescale of this transition
depends on the accretion rate so our emphasis remains qualitative, but
we could imagine the following scenario.  Because the transition
involves going from an FRII to an FRI state, a jet morphology might
appear in which further away from the black hole engine is the
presence of the relic, well-collimated FRII double-sided jet, whereas
closer to the black hole appears the younger double-sided FRI jet,
such as in 3C\,288 (Bridle et al. 1989; Lal et al. in preparation).
Alternatively, the likelihood is that of a spin-flip without
transition through zero spin.  The spin value of that flip would produce
an even larger break in the boundary between FRIIs and FRIs.  This
transition or gap in the efficiency of the engine near zero spin or
due to a spin-flip is important in that it provides the possibility of
a sharp break between states that have high and low BP, suggesting a
natural location for a break between collimated and uncollimated or
less collimated jets.

We also emphasize the fact that our model is founded on the
prescription of prolonged accretion as opposed to chaotic accretion
scenarios (Volonteri, Sikora \& Lasota 2007; Berti \& Volonteri 2008;
King, Pringle \& Hoffman 2008) so that most of the mass is provided by
an accretion disk with fixed orientation (Miller 2002), and thus, as
pointed out in more detail in the discussion section, we expect black
holes in all galaxies to evolve toward highly spinning prograde values
as a result.  The chaotic accretion scenario is usually invoked to
produce low spins in radio-quiet spiral galaxies under the assumption
that the spin paradigm operates; but, given that radio-quiet AGN are
not low-spinning black hole accretion systems in the gap paradigm, no
such mechanism need be invoked (as pointed out by Berti \& Volonteri
(2008), the spin of $0.99$ claimed by Brenneman \& Reynolds for
MCG-06-30-15, is very unlikely in the chaotic accretion scenario.)  In
other words, {\em accretion in the gap paradigm simply spins all black
holes up in the prograde direction in all galaxies, ellipticals as
well as spirals}.  Accordingly, the observed distribution of powerful
radio-loud galaxies reflects the ability of mergers to produce
retrograde accretion flows in our model, which implies that mergers
are constrained to produce a peak of such occurrences at $z\approx2$,
a distribution that dies off at higher and lower $z$.

\label{cosmo_evolution}

\subsection{FRI/II Dichotomy}

The FRI/FRII dichotomy between low radio-luminosity, poorly collimated
radio jets and high radio-luminosity, highly collimated radio jets,
has been explained according to two different physical scenarios.  The
first assumes that external environmental factors influence the jet
structure (De Young 1993; Laing 1994; Bicknell 1995; Kaiser \&
Alexander 1997; Gopal-Krishna \& Wiita 2000), while the second
attributes the differences in morphology to parameters associated with
the jet production mechanism itself (Rees 1982; Baum et al 1995;
Reynolds et al 1996; Meier 1999, 2001; Ghisellini \& Celotti 2001;
Marchesini et al. 2004).

\begin{figure*}
\centerline{\includegraphics[angle=-0,scale=0.60]{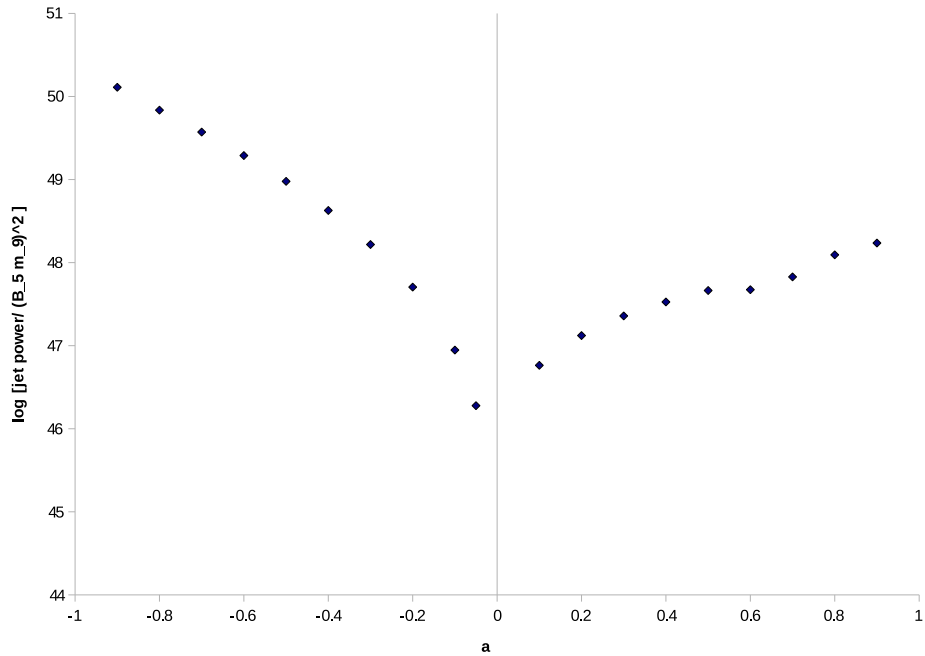}}
\caption{The overall jet power assuming BZ and BP operate in tandem
 via equation \ref{jet_power}.  $B_{5}$ is magnetic field in units of
 $10^{5}$ Gauss and $m$ is the ratio of black hole mass to $10^{9}$
 solar masses (i.e. for a $10^{5}$ Gauss field, a $10^{9}$ solar mass
 black hole, and a retrograde spin of -0.9, the power is slightly
 above $10^{50}$ erg/s).  The value of $\delta$ is at a conservative
 estimate of about $2.5$ but could be a magnitude or more higher.
 Numerical simulations of jets are needed to determine this value
 precisely.  We emphasize that the discussion in section
 \ref{Conclusions} implies that the above jet power dependence becomes
 irrelevant at high prograde spin in HERGs since the efficiency shifts
 from jets to disks.  }\label{Ledlow}
\end{figure*}

\begin{figure*}
\centerline{\includegraphics[angle=-0,scale=0.80]{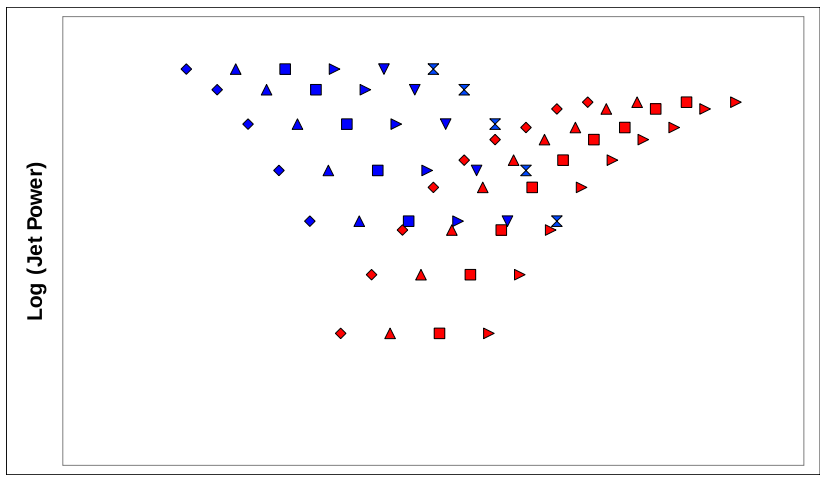}}
\caption{Going from jet power vs spin state to the Ledlow-Owen
diagram.  The x-axis can have black hole mass, time or galaxy optical
luminosity depending on the relative stretching between retrograde and
prograde regimes.  FRII diamonds (blue) evolve into FRI diamonds
(red); FRII triangles (blue) evolve into FRI triangles (red); FRII
squares (blue) evolve into FRI squares (red) etc.  We have combined
multiple theoretical evolutionary paths and have focused on the
transition region between FRIIs (blue) and FRIs (red) at lower power
(i.e. we have focused on the region of Figure \ref{Ledlow} where spin
varies between $-0.2$ and $0.2$ and the y-axis is below unit $47$). }
\label{Owen-Ledlow} \end{figure*}

Our model weighs in on this issue by naturally reproducing an
FRI/FRII, Ledlow-Owen-like diagram as we illustrate below.  Figure
\ref{Ledlow} shows the power in the jet as a function of spin state.
According to our scenario, FRIIs are retrograde accretion systems and
so are located on the left side of the break centered at zero spin.
FRIs, instead, are modeled as prograde spin states so are located on
the right-hand-side of the break.  The time flow in our model runs
from retrograde accretion toward prograde accretion.  Therefore, the
x-axis can be replaced by time.  Such a jet power vs time diagram
serves the purpose of illustrating the cosmological evolution of a
single initial FRII HERG.  In attempting to recover a Ledlow-Owen-like
diagram from our model, we must consider a combination of the time
evolution of multiple initial FRII HERGs.  The simplest way to do this
is to imagine a series of FRII HERGs forming at different times,
producing a separation between otherwise similar paths.  In other
words, we have combined, side-by-side, paths like those of
Fig. \ref{Ledlow} and simply shifted them along the x-axis, producing
multiple plots of translated but equal configurations.  Since
supermassive black hole mass increases in time due to accretion, from
a qualitative perspective, similar information appears on both a time
and black hole mass plot despite a relative stretching between such
diagrams due to short lifetimes and smaller mass increase during the
retrograde phase compared to the prograde one.  In other words, one
can produce a qualitatively valid diagram by exchanging time with
black hole mass.  Finally, via the black hole mass-galaxy luminosity
relation, one can exchange black hole mass with optical galaxy
luminosity on the x-axis.  These steps constitute a qualitative,
zeroth-order attempt, at illustrating the consequences of our model on
the evolution of a $\it{family}$ of powerful radio-loud objects.  The
result is Figure \ref{Owen-Ledlow}.  From a qualitative perspective,
then, the Ledlow-Owen diagram marks a division between objects that
are young and powerful and others that are old and weak.  Although
that is the general trend, near the transition region of Figure
\ref{Ledlow}, we can appreciate the existence of FRIIs that have
weaker jets compared to FRIs further up beyond the break in the
prograde regime.

The presence of a large population of FRIIs (blue) in the lower part
of Figure \ref{Owen-Ledlow} as opposed to the actual Owen-Ledlow
diagram is simplistic in two ways.  First, the diagram fails to
capture the fact that the lower power FRII engines are less collimated
than their more powerful counterparts suggesting a less superficial
classification in which ``FRII-like'' or ``blue-like'' objects
populate the lower regions of Figure \ref{Owen-Ledlow}.  Second, we
must also consider the effects of the external environment.  As galaxy
luminosity increases to the right on the x-axis, the galactic
environment is more prone to turning these otherwise low power FRIIs
into FRIs.  Applying the same argument to the FRIIs (blue) that are
mixed in with the FRIs (red) further up but to the right in the
diagram of Figure \ref{Owen-Ledlow} (i.e. where the objects are living
in more dense stellar environments compared to their counterparts at
equal jet power further back along the x-axis) suggests that
environment might be responsible for turning some of them into FRIs.
In other words, the further to the right we move on the diagram, the
further up in the vertical direction we expect to replace FRIIs with
FRIs.  Because, to re-emphasize, our theoretical paths of Figure
\ref{Owen-Ledlow} are based solely on the properties of the jet
engine, it is not surprising that it is precisely that which we are
not modeling (i.e. environment) that is required to make theory
compatible with observation.  In other words, the Fanaroff-Riley
dichotomy remains a function of jet power (governed by black hole
spin) and environmental factors.  We recognize that producing Figure
\ref{Owen-Ledlow} does not yield additional predictive power.
However, we suggest that the simple compatibility at even a
superficial level between the gap paradigm and the actual Owen-Ledlow
diagram is statistically non-trivial.
\label{Fanaroff-Riley}

\subsection{Fe K$\alpha$ Lines and Disk Winds}

One interpretation of the narrowness of iron lines in HERGs compared
to radio-quiet AGN is that the innermost stable circular orbit is
further away from the black hole for higher retrograde accretion
systems (we will suggest that true radio-quiet AGN are maximally
spinning prograde systems).  As far as reflection features in the
inner disk regions go, even mild relativistic motion in coronal
material above an accretion disk can reduce the reprocessed radiation
from the disk (Beloborodov 1999).  At velocities of just under $0.3c$,
fluorescent line emission is effectively washed out (Reynolds \&
Fabian 1997).  Therefore, the weakness of reflection features in HERGs
may simply be the natural consequence of highly relativistic motion
away from the disk in the inner regions where both BZ and BP conspire
to produce relativistic jets while their narrowness indicates disk
inner edges at larger radii.

The largeness of the gap region that produces the powerful, collimated
jets of the FRII HERGs is also associated with an absence of
inner-disk regions that are close to the black hole where
gravitational potential energy can be reprocessed in the disk.  As
mentioned above, the absence of this reservoir of gravitational energy
is compatible with a decrease or absence of disk winds further out in
the disk.  FRI LERGs, on the other hand, also fail to produce disk
winds but due to the fact that the flow is advective dominated and not
thin-disk-like.  FRI HERGs, according to our model, should possess
disk winds because the flow is thin-disk-like and the gap region is
smaller so the reservoir of reprocessable gravitational energy near
the black hole for use further out in the disk is comparatively
larger.  We will argue, however, that as the spin increases in the
prograde direction, the increase in disk winds resulting from smaller
ISCO radii, inhibit jets, so that FRI HERGs fails to be appropriate
nomenclature once jets are absent.  As a result, FRI HERGs applies
only to lower prograde HERGs.

The fact that FRI LERGs are located on the right-hand side of Figure
\ref{Ledlow} implies that such objects should display a weak
dependence on black hole spin since the power is roughly a flat
function of spin for a range of intermediate spins.  As pointed out
(Garofalo 2009b), this alleviates the fine-tuning issue of Nemmen et
al (2007) and Benson \& Babul (2009).  In addition, the retrograde or
FRII region involves a much steeper function of black hole spin so the
dependence on accretion rate should not be the same as it is for the
FRIs.  In other words, our model removes the need for even higher
accretion rates in order to model the most powerful FRIIs,
i.e. radio-loud quasars.  It might be worth emphasizing that although
jet power in the BZ effect does not come from the accretion power per
se, the magnetic field which allows extraction of black hole
rotational energy depends on the accretion rate in such a way that for
fixed black hole spin, BZ power depends linearly on accretion rate in
ADAFs.  We also point out that retrograde accretion states are
short-lived.  From purely accretion considerations, a highly-spinning
retrograde black hole will be spun down to zero spin when the black
hole has accreted about 0.2 of its original mass (Moderski \& Sikora
1996).  The prograde evolution regime, on the other hand, is slower
not only because the black hole must reach a mass of $\sim 2.5M$ where
$M$ is the original black hole mass (Volonteri, Sikora \& Lasota
2007), but also because the FRI regime is characterized by low angular
momentum black hole-feeding in radiatively inefficient accretion.
While FRIs have lifetimes reaching $\sim 10^{8} years$, FRII lifetimes
are about a factor of ten smaller at a few times ($10^{6}-10^{7}$)
years (O'Dea et al. 2009).
\label{Fe_alpha}

\section{Discussion and Conclusion}

We presented a framework for the cosmological evolution of radio-loud
galaxies in which the gap region between accretion disks and black
holes is key.  In this gap paradigm, jet production is most effective
when black hole magnetospheres conspire with their large gap regions
to produce strong black hole-threading magnetic fields that are
geometrically favorable to both the BZ and BP effects.  As time
evolution through accretion spins the black hole up in the prograde
direction and the gap region size decreases, the interaction of
rotating black holes with their magnetospheres becomes inefficient and
jets weaken, become less collimated and if accretion spins the black
hole up to high prograde values, likely fails entirely to produce
jets.  Maximally spinning prograde black hole magnetospheres that are
the result of cosmological evolution via accretion, thus, may be
systems that have now become inactive.  The radio-quiet quasars formed
in ellipticals may, instead, involve the population of post-merger
systems characterized by high accretion onto spinning black holes in a
prograde configuration.  According to our scenario, the origin of the
radio-loud/radio-quiet dichotomy may lie in the existence of two types
of engine efficiencies.  On the one hand, retrograde black holes
conspiring with their magnetized flows are highly efficient in
producing non-thermal, jet outflows, but shorter-lived configurations.
These engines produce the observed radio-loud population.  On the
other hand, prograde, high accretion systems with small gap regions,
while incapable of producing powerful jets, are capable of tapping
into the gravitational potential energy of their inner accretion flow,
making them highly efficient in producing strong thermal emission.
The reprocessed gravitational potential energy of the inner regions
near the rapidly spinning black hole is fundamental in producing the
disk winds from larger disk radii.  It has recently been shown that
GRS 1915+105, a galactic microquasar, exhibits a soft state in which
the radiation field drives a hot wind off the accretion disk carrying
enough mass to halt the flow of matter into the radio jet (Neilsen \&
Lee 2009), a picture that is underscored by theoretical work (Meier
1996) and for which observation is mounting in other systems as well
(Blum et al. 2010).  This scenario suggests that most FRI HERGs are
systems in which the disk winds eventually inhibit jet production
making such objects part of the radio-quiet population, thereby
explaining the rarity of FRI HERGs.  The handful of observed FRI HERGs
are then systems where the black hole spin is small, producing
relatively larger gap regions, which decreases the production of winds
further out from the disk and allows the presence of an FRI jet.  In
other words, in the cosmological evolution of radio-loud objects, FRI
HERGs are objects that are transitioning away from strong jet
production and weak winds, to a scenario involving weaker jets and
stronger winds but that have not reached the high prograde spin states
that would produce strongest winds and weakest or no jets.  We
highlight the compatibility of this scenario with the existence of FRI
type radio quasars (Heywood et al 2007) as pointed out to us by David
L. Meier.  The fact that accretion states transition from HERGs to
LERGs, however, allows such systems to remain radio-loud as they
increase their prograde spins as a result of the failure of ADAFs to
produce the strong jet-inhibiting disk winds.  Thus, spinning black
holes produce both FRII HERGs and radio-quiet quasars depending on
retrograde vs prograde accretion.

If mergers are responsible for both the high accretion rates and the
occurrence of retrograde flows, it does not surprise that spiral
galaxies are not competitive in their radio-loudness, and less
accretion powered than the post-merger ellipticals.  If retrograde
accretion is formed in mergers, it also does not suprise that spirals
are predominantly radio-quiet.  In fact, only the low prograde spin
configurations would lead to radio-loudness.  Given that higher
prograde accretion systems surrounded by radiatively efficient thin
disks have smallest gap regions and the strongest jet-inhibiting
winds, higher prograde spin systems in spirals should be radio-quiet;
and, therefore, maximally spinning prograde configurations should be
the most radio-quiet sources.  In short, prograde vs retrograde,
radiatively efficient accretion flows, produce the
radio-quiet/radio-loud dichotomy.

And finally, we conclude our discussion with a glance at the
implications for microquasars.  Because such objects are prograde
accretion systems, their accretion state changes produce the
microquasar counterpart to AGN HERGs and LERGs, or more appropriately,
a transition between high prograde spin, high excitation accretion
states (or radiatively efficient accretion states) to high prograde
spin, low excitation accretion states (or radiatively inefficient
accretion states).  If the parallel with the FRII feedback phase in
AGN is valid, these high prograde spin, high excitation accretion
states in microquasars produce strong disk winds that expel the gas
(Neilsen \& Lee 2009), thereby producing an ADAF phase during which
disk winds drop and jets are no longer hindered while the system
enters the radio dominated hard state.  Eventually, the accretion flow
from the companion allows the system to re-enter the soft or
radiatively efficient accretion state.  The extension of our model to
microquasars suggests that the approach to the high excitation
accretion (soft) state from the low excitation accretion (hard) state,
involves the brief combined presence of a BZ component and a BP-like
component produced by the onset of inner disk winds.  In other words,
there is a short time during which the microquasar mimicks the
conditions in FRII HERGs, thereby producing a powerful, collimated
jet.  Note that no spin change occurs.  This suggests that
microquasars undergo state transitions that take them from high
prograde spinning high excitation accretion states (soft states) to
high prograde spinning low excitation accretion states (hard states)
and back, and that during the low excitation to high excitation
transition, the physical conditions of an FRII HERG are briefly
produced.

We summarize our main conclusions as follows:

\begin{enumerate}
\item FRIIs have retrograde spin while FRIs have prograde spin
\item FRII HERGs evolve toward FRI LERGs
\item FRIIs have larger jet efficiencies while FRIs have lower jet
efficiencies
\item High prograde spin systems in HERG states have large disk
efficiencies (and thus strong disk winds) while LERG states and
retrograde accretion states have lower disk efficiencies (and thus
lower disk winds)
\item Highly efficient jet engines are highly inefficient disk engines
and viceversa
\item FRI HERGs have low prograde spin
\item Radio quiet AGN are high spinning prograde, radiatively
efficient systems
\item Maximally spinning prograde, radiatively efficient systems, are
the most radio quiet AGN
\item Spiral galaxies involve prograde accretion
\item Spiral galaxies should be more radio loud when the spin is lower
and more radio quiet, quasar-like, when the spin is high
\item A radiatively inefficient to radiatively efficient transition,
in high prograde systems, produces a short-lived, powerful, collimated
jet as observed in microquasars. 
\end{enumerate}

We conclude by highlighting the fundamental role or impact of general
relativity in galaxy evolution in this model.  The validity of our
paradigm suggests that aspects of the spacetime metric that are
assumed to govern physics in regions provincially relegated to the
near black hole, are in fact dominant on much larger scales, to the
extent that galaxy morphology, energetics, and evolution, are tightly
linked to the details of strong-field general relativistic effects.
Whereas a Newtonian treatment of spacetime is successful just a few
gravitational radii from the black hole in the sense that relativistic
corrections are negligible there, the scale of influence of black hole
spin that our paradigm forces us to grapple with is daunting, with up
to more than eight orders of magnitude beyond its local sphere of
influence.  It is the details of tiny regions of highly curved
spacetime that have the greatest effect on the large-scale properties
of galaxies.

\label{Conclusions}

\section{acknowledgments}

D.G. thanks David L. Meier for discussion on the FRI/FRII division,
the physics of jet launching, propagation and collimation, and for
commenting on the draft as a whole, Peter Polko for stimulating ideas
on the effects of BP jets on accretion disk coronae, Christopher
S. Reynolds for issues concerning broad iron lines and jet energetics
and Cole Miller and Marta Volonteri for discussion on retrograde
accretion.  We acknowledge the role of and thank the three referees
and three editors that were involved in this project.  The research
described in this paper was carried out at the Jet Propulsion
Laboratory, California Institute of Technology, under a contract with
the National Aeronautics and Space Administration.  D.G. is supported
by the NASA Postdoctoral Program at NASA JPL administered by Oak Ridge
Associated Universities through contract with NASA.

\section*{References}


\noindent Allen S.W. et al, MNRAS, 2006, 372, 21

\noindent Antonuccio-Delogu V. \& Silk J., 2010, MNRAS, submitted

\noindent Arshakian, T., Beck, R., Krause, M., Sokoloff, D., \&
Stepanov, R., Proceedings to Panoramic Radio Astronomy, Groningen,
2009

\noindent Barnes J.E.\& Hernquist, L. 1991, ApJ, 370, L65 

\noindent Baum S.S., Zirbel E.L. \& O'Dea C.P., 1995, ApJ, 451, 88

\noindent Beloborodov, A.M., 1999, ApJ, 510, L123

\noindent Benson A.J. \& Babul A., 2009, MNRAS, 397, 1302

\noindent Bicknell G.V., 1995, ApJS, 101, 29 

\noindent Blandford, R. D., \& Payne, D. G. 1982, MNRAS, 199, 883

\noindent Blandford, R. D., \& Znajek, R. L. 1977, MNRAS, 179, 433

\noindent Blandford, R.D., in Active Galactic Nuclei,
ed. T.J.-L. Courvoisier \& M. Mayor, 161-275

\noindent Blum J.L. et al, 2010, ApJ, in press

\noindent Bogovalov, S. \& Tsinganos, K. 2005, MNRAS, 357, 918

\noindent Brenneman, L., \& Reynolds, C.S., 2006, ApJ, 652, 1028

\noindent Bridle A.H. et al, 1989, AJ, 97, 674

\noindent Cao X., 2003, ApJ, 599, 147

\noindent Daly, R.A., ApJ, 2009, 691, L72

\noindent Daly, R.A., ApJ, 2009, 696, L32

\noindent De Young D.S., 1993, ApJ, 405, L13 

\noindent Donato D., Sambruna, R.M., Gliozzi, M, 2004, ApJ, 617, 915

\noindent Evans, D., 2004, ApJ, 612, 786 

\noindent Evans, D., et al, 2006, ApJ, 642, 96

\noindent Fabian, A.C. et al, Nature, 2009, 459, 540

\noindent Fanaroff, B.L.\& Riley, J.M., 1974, MNRAS, 167, 31P

\noindent Ferrarese, L., \& Merritt, D., 2000, ApJ, 539, L9

\noindent Garofalo, D., 2009, ApJ, 699, L52 (a)

\noindent Garofalo, D., 2009, ApJ, 699, 400 (b)

\noindent Gebhardt, K., et al. 2000, ApJ, 539, L13

\noindent Ghisellini G., Celotti A., 2001, A\&A, 379, L1 

\noindent Gopal-Krishna \& Wiita P.J, 2000, A\&A, 363, 507

\noindent Grandi P., Urry C.M., Maraschi L., 2002, NewAR, 46, 221

\noindent Kormendy, J., \& Richstone, D., 1995, ARA\&A, 33, 581

\noindent Hardee P.E \& Hughes P.A., 2003, ApJ, 583, 116

\noindent Hardcastle M.J., Evans D.A. \& Croston J.H., 2006, MNRAS, 370, 1893

\noindent Hardcastle M.J., Evans D.A. \& Croston J.H., 2007, MNRAS, 376, 1849

\noindent Heywood I., Blundell K.M.\& Rawlings S., 2007, MNRAS, 381, 1093

\noindent Hughes, S.A. \& Blandford, R.D., 2003, ApJ, 585, L101

\noindent Kaiser C.R. \& Alexander P., MNRAS, 1997, 286, 215 

\noindent King A.R., Lubow S.H., Ogilvie G.I. \& Pringle J.E., 2005,
MNRAS, 363, 49

\noindent King, A.R., Pringle,J.E. \& Hofmann, J.A., 2008, MNRAS, 385, 1621 

\noindent Kormendy, J., \& Richstone, D., 1995, ARA\&A, 33, 581

\noindent Kuncic, Z. \& Bicknell, G.V., 2004, ApJ, 616, 669

\noindent Kuncic, Z. \& Bicknell, G.V., 2007, Ap\&SS,311, 127 

\noindent Laing R.A., 1994, ASPC, 54, 227

\noindent Lal  et al, in preparation

\noindent Livio, M., Ogilvie, G.I. \& Pringle, J.E., 1999, ApJ, 512, 100

\noindent Magorrian, J., et al. 1998, AJ, 115, 2285

\noindent Mangalam A., Gopal-Krishna \& Wiita P.J., 2009, MNRAS, 397, 2216

\noindent Marchesini D., Celotti A. \& Ferrarese L., 2004, MNRAS, 351, 733 

\noindent Marconi, A., \& Hunt, L.K., 2003, ApJ, 589, L21

\noindent McKinney J.C. \& Narayan R. 2007, MNRAS, 375, 513

\noindent McKinney J.C. \& Blandford R.D., 2009, MNRAS, 394, L126

\noindent McKinney J.C. \& Gammie C.F., 2004, ApJ, 611, 977

\noindent McNamara B.R., et al,2009, ApJ, 698, 594

\noindent Menon, G. \& Dermer, C.D., 2005, ApJ, 635, 1197

\noindent Meier D. L., 1999, ApJ, 522, 753

\noindent Meier D.L. et al., Science 291 (2001), 84

\noindent Meier D.L., 1996, ApJ, 459, 185

\noindent Miller, M.C., 2002, ApJ, 581, 438

\noindent Moderski R.\& Sikora M., 1996, A\&AS, 120, C591

\noindent Moderski R., Sikora M.\& Lasota J.-P., MNRAS, 1998, 301, 142

\noindent Nakamura, M. \& Meier, D.L., 2004, ApJ, 617, 123

\noindent McClintock, J.E., Shafee, R., Narayan, R., Remillard, R.A.,
Davis, S.W.\& Xin-Li, L., ApJ, 652, 518

\noindent Narayan, R. \& Yi I., 1995, ApJ, 452, 710

\noindent Neilsen J. \& Lee J., Nature, 458, 481 

\noindent Neemen R.S. et al, 2007, MNRAS, 377, 1652

\noindent Nesvadba N.P.H, Lehnert M.D., De Breuck C., Gilbert A. \&
van Breugel W., 2008, A\&A, 491, 407

\noindent Novikov I. \& Thorne K., 1973, Black Holes (Les astres occlus), p. 343

\noindent O'Dea et al, 2009, A\&A, 494, 471

\noindent Ogle P., Whysong D. \& Antonucci R., 2006, ApJ, 647, 161

\noindent Rees M.J., 1982, IAUS, 97, 211 

\noindent Reeves J.N.\& Turner M.J.L., 2000, MNRAS, 316, 234

\noindent Reeves J.N., Sambruna R.M., Braito V. \& Eracleous
M., 2009, ApJ, 702, L187

\noindent Reynolds C.S. et al, 1996, MNRAS, 283, L111

\noindent Reynolds C.S. \& Fabian A.C. 1997, MNRAS, 290, L1

\noindent Reynolds C.S., Garofalo D.\& Begelman M.C., 2006, ApJ, 651, 1023

\noindent Sambruna et al, 2009, ApJ, 700, 1473


\noindent Smolcic V., 2009, ApJ, 699, L43

\noindent Smolcic V. et al, 2009, ApJ, 696, 24

\noindent Tchekhovskoy A., McKinney J.C. \& Narayan R. 2009, ApJ, 699, 1789

\noindent Tchekhovskoy A., Narayan R \& McKinney J.C. 2010, ApJ, in press

\noindent Thorne K.S., Price R.H. \& Macdonald D.A., 1986, Black
Holes: The Membrane Paradigm (New Haven: Yale Univ. Press)

\noindent Tremaine S., et al 2002, ApJ, 574, 740

\noindent Volonteri M., Sikora M., \& Lasota J-P, 2007, ApJ, 667, 704

\noindent Wilms J. et al, 2001, MNRAS, 328, L27

\noindent Wilson A.S. \& Colbert E.J.M, 1995, ApJ, 438, 62

\noindent Zoghbi A. et al, 2010, MNRAS, 401, 2419


\end{document}